\documentclass[12pt,a4paper]{article}
\usepackage[dvips]{graphicx}
\begin{document}
\title{KNO scaling in processes of electron-positron annihilation
to hadrons}
\author{V.\,A.\,Abramovsky\/\thanks{ava@novsu.ac.ru}, N.\,V.\,Radchenko\\
        Novgorod State University, B. S.-Peterburgskaya Street 41,\\
        Novgorod the Great, Russia, 173003}
\date{}
\maketitle

\begin{abstract}The charged particles multiplicity distribution in the
KNO form is discussed in processes of $e^+e^-$ annihilation at
energies $\sqrt{s}$ 14 -- 206.2 GeV. The experimental data are
compared to data, obtained with Monte Carlo simulation in PYTHIA
in the Lund quark string model. It is shown, that both
experimental and simulated data are described by the same
distribution function in the KNO form. It is shown, that the KNO
scaling is  consequence of quark string hadronization dynamics in
the Lund string model.\end{abstract}

Since the time of publication of Koba, Nielsen and Olesen
paper~\cite{bib:1} (and also of Polyakov paper~\cite{bib:2}, which
had appeared 2 years earlier) it became clear that the scaling
invariance of multiplicity distribution function at high energies,
the so-called KNO scaling, gives an important information on
character and dynamics of multiple hadron production.

The simplest processes are the processes of hadron production in
$e^+e^-$ annihilation, in which hadrons are produced in strings of
color field that have quark and antiquark at the endpoints, the
so-called quark strings~\cite{bib:3}.

The advantage of these processes both from theoretic and
experimental points of view is that there is no need to take into
account the parton distributions in the initial state, unlike in
deep inelastic scattering. It allows one to avoid uncertainties
connected with the phenomenological character of these parton
distributions. Moreover, since there are no hadrons in the initial
state, the produced secondary hadrons completely represent the
dynamics of quark strings decay (or any other hadronization
mechanism).

In the present study we will show that the charged particles
multiplicity distribution functions in processes of $e^+e^-$
annihilation to hadrons, obtained experimentally to the total
center-of-mass energy 206.2 GeV, satisfy the KNO scaling. We will
also show that to this energy contribution to the dynamics of the
process is given by the only one quark string, and  this is
exactly what determines the presence of the KNO scaling.

We consider the process of $e^+e^-$ annihailation at 6 values of
total center-of-mass energy -- $\sqrt{s}$ from 14 to 206.2 GeV.
Wide coverage of the experimental data and  sufficient statistical
provision of them were criteria for the choice. At the energy 14
GeV the number of events is 2\,704 (TASSO)~\cite{bib:4}, at 29 ÃýÂ
-- 29\,649 (HRS)~\cite{bib:5}, at 34.8 ÃýÂ -- 52\,832
(TASSO)~\cite{bib:4}, at 91.2 ÃýÂ -- 248\,100 (L3)~\cite{bib:6},
at 188.6 ÃýÂ -- 4\,479 (L3)~\cite{bib:6}, at 206.2 ÃýÂ -- 4\,146
(L3)~\cite{bib:6}.

The distribution function $<\!\!n\!\!>P_n$ is shown in Fig.1,
where $P_n$ is the probability of production of $n$ charged
particles, and $<\!\!n\!\!>$ is the mean multiplicity of the
charged particles at given energy. The  ordinate axis in Fig.1a is
given in linear scale in order to demonstrate coincidence in the
main part of the peak. We think that slight discrepancy near the
maximum of the distribution is related firstly to the experimental
errors, and secondly, but this is the most important in our sight,
to transition from the discrete multiplicity distribution to the
continuous distribution. Maximum in the discrete distribution is
absolutely not required to coincide with true maximum. The
behavior of the function in tails of the distribution that
correspond to low and high multiplicities is shown in Fig.1b,
where the ordinate axis is given in logarithmic scale.

The points at all energies are described by the same distribution
function within the errors. In order to more explicitly
demonstrate this, the distribution functions $<\!\!n\!\!>P_n$ of
the minimal (14 GeV) and the maximal (206.2 GeV) of the discussed
here energies are given in Fig.2. Visually all points lie on one
curve both for linear scale (Fig.2a) and for logarithmic scale
(Fig.2b).

Hence, we can accept that in the energy range from 14 GeV to 206.2
GeV, the distribution function $<\!\!n\!\!>P_n$ has the KNO form
$$
<\!\!n\!\!>P_n=\Psi\left(\frac{n}{<\!\!n\!\!>}\right).
$$
It results from the KNO scaling that a ratio $<\!\!n\!\!>/D$,
where $D=\sqrt{<\!\!n^2\!\!>-<\!\!n\!\!>^2}$ is a dispersion, does
not depend on the energy. These data are given in Table~1.
Moreover, the approximate constancy (within the errors) of the
experimental values of the highest multiplicity distribution
moments, $C_l=<\!\!n^l\!\!>/<\!\!n\!\!>^l$, $l=3,4,5$, also
supports the presence of the KNO scaling (Table~2).

The fact of the experimental points lying on the KNO curve in
different energy ranges was presented in many papers, in
particular, in ~\cite{bib:4} ($\sqrt{s}$ 3.6 -- 43.6
GeV),~\cite{bib:5} ($\sqrt{s}$  14 -- 34.5 GeV),~\cite{bib:7}
($\sqrt{s}$  29 -- 91.2 GeV),~\cite{bib:8} ($\sqrt{s}$  14 -- 91.0
GeV). At the energies higher than 91.2 GeV the curves are not
considered, only the ratio $<\!\!n\!\!>/D$ is
presented~\cite{bib:9} ($\sqrt{s}$  14 -- 200 GeV). In the
papers~\cite{bib:4},~\cite{bib:8},~\cite{bib:10} it was emphasized
that the KNO scaling in the Lund string model holds only
approximate and occurs as the result of an accidental combination
of several effects. Particulary, it was stated that at low
energies energy-momentum conservation law leads to narrowing of
the multiplicity distributions. On the other hand, at higher
energies production of $b$ quarks and hard gluon emission increase
the dispersion of the multiplicity distribution, and together
these two effects lead to the appearance of the approximate
KNO-scaling.

The results presented in Fig.1 and 2 show good coincidence in so
large energy range which, in our opinion, can not be accidental.
In what follows we will state the arguments, that the
experimentally observed in the energy range from 14 to 206.2 GeV
KNO scaling is the consequence of the well defined hadronization
dinamics.

All theoretical results, including the first work on the KNO, are
obtained in far on energy asymptotic for which the value of
$1/\ln\sqrt{s}$ is neglecting small. Moreover, it seems to
impossible to get the regularities of such complicated event with
many degrees of freedom as multiple hadron production from the
first principles of QCD. So we have to use more phenomenological
and model approaches. We have chosen the Lund string model and its
realization in Monte Carlo generator PYTHIA~\cite{bib:13} that
provides good description of lots of the experimental data. It
should be noted that PYTHIA does not take into account the KNO
scaling phenomenologically~\cite{bib:14}.

Electron-positron annihilation to hadrons was simulated with
PYTHIA at every discussed energy from 14 to 206.2 GeV. The decay
process of $\gamma^\ast\left(Z^0\right)$ was considered without
the initial state radiation, values of the other parameters were
set by default. There were generated one million events for every
energy, and the charged particles multiplicity distributions were
analyzed, the obtained results are presented in Fig.3.

The distribution function $<\!\!n\!\!>P_n$ has the KNO form at
every energy. The simulated values of mean multiplicity
$<\!\!n\!\!>$, dispersion $D$ and ratio $<\!\!n\!\!>/D$ are
presented in Table~1. All these values are in good agreement with
the experimental data. The values of moments $C_3$, $C_4$, $C_5$
are presented in Table~2, they are also coincide with the
experimental values. If one combines Fig.1 and Fig.3 then we get
good visual agreement of the experimental and simulated data. In
what follows we will consider that both distributions are
described by the same KNO function.

It was noted before that the main process of hadronization in
$e^+e^-$ annihilation is decay of quark string of color field.
This string appears between flying quark-antiquark pair produced
from $\gamma^\ast$ or $Z^0$. The mean number of hadrons produced
from decay of the quark string is associated with the string
length in rapidity space, $l$. We will not concretize this
dependence here. The quark-antiquark pair produced from
$\gamma^\ast\left(Z^0\right)$ can produce the bremsstrahlung
gluons which in their turn can produce additional quark-antiquark
pairs. The other bremsstrahlung gluons are assimilated by the
quark-antiquark pairs producing in the string decay process.
Therefore, besides one quark string, there can be several strings,
each of which will pass into hadrons.

The distributions of number of strings obtained from the same
PYTHIA simulated events for every energy are presented in Table~3.
As the string length in rapidity space we take the value
$$
l=\ln\frac{(P_i+P_j)^2}{(m_i+m_j)^2},
$$
where $P_i$, $P_j$ are the four-momentum of the quarks that hold
the string endpoints, $m_i$, $m_j$ are the constituent masses of
these quarks. The mean summarized string length, the mean charged
particles multiplicity and the weight of the corresponding events
for events with one string, two strings and so on are presented.

One can imagine the following hadronization picture  in $e^+e^-$
hadronization. At the low energies there is only one quark string
between  quark-antiquark pair produced from
$\gamma^\ast\left(Z^0\right)$ (Fig.4a). Production of an
additional pair is suppressed, its probability is only few
percents. With the growing energy the weight of the additional
quark-antiquark pair is increasing. This pair can lead to
splitting of one quark string to two strings (Fig.4b). Since the
additional quark and antiquark have approximately the same momenta
the endpoints of string from this pair will not be far from each
other in rapidity space. So we can consider these two strings as
one string without splitting and with the effective length equal
to the sum of the lengths of two strings.

As it is seen from Table~3, the lengths of one and two strings for
every energy are fully placed in the allowed rapidity phase space,
which is defined as $\ln[s/4(m_{ud})^2$], where $m_{ud}=0.33$ GeV
is the constituent mass of $u$ and $d$ quarks. For the energies
under consideration it gives 6.10; 7.56; 7.93; 9.86; 11.30 and
11.50 accordingly. The smaller length of one quark string is bound
with the radiation of several bremsstrahlung gluons by the
quark-antiquark pair produced from $\gamma^\ast\left(Z^0\right)$,
thus decreasing the energies of quark and antiquark. The precise
KNO scaling by its definition must not depend on the energy of
flying quarks, i.e. it must not depend on the string length.
Therefore, both for one and two quark strings the KNO distribution
function $\Psi(n/<\!\!n\!\!>)$  will have the same form; resulting
function will present the sum of these two identical
distributions. Violation of the KNO appears only when strings
overlap, as it was shown in~\cite{bib:15}. But  overlapping of the
strings is possible only when two or more additional
quark-antiquark pairs appear, i.e. when there are three or more
quark strings (Fig.4c). The summarized weight of these processes
is only 3\% at 206.2 GeV (Table~3). Therefore the KNO scaling is
not violated for all energies considered in this paper. In order
to check the assumption of equivalence of two quark strings to one
effective string, the charged particle multiplicity distribution
functions in the KNO form were obtained separately for events with
only one quark string and for events with two quark strings. The
KNO distribution function for events with one string is shown in
Fig.5 (the  ordinate axis is given in linear scale). The
multiplicity distribution for events with two quark strings
practically entirely coincides with the case of one string. At
that time the KNO scaling is strongly violated when one draws the
charged particles distribution function in the KNO form for events
with three quark strings.

If one combines the distributions $\Psi(n/<\!\!n\!\!>)$ given in
Fig.1a, Fig.3a and Fig.5, which are drawn in the same scale, then
we obtain the completely coincidence of all distributions.

Hence, we can state that the KNO scaling obtained for all
experimental data at the energy range 14 -- 206.2 GeV is not
accidental. We have based on the Lund string model of quark string
which, as one can expect, describes the main characteristics of
the multiple hadron production well.

Authors are grateful for A.\,V.~Dmitriev and N.\,V.~Prikhod'ko for
useful discussions.

This work was partially supported by RFBR grant 07-07-96410-r
center-a.

\begin{figure}
\includegraphics[height=115mm,width=155mm]{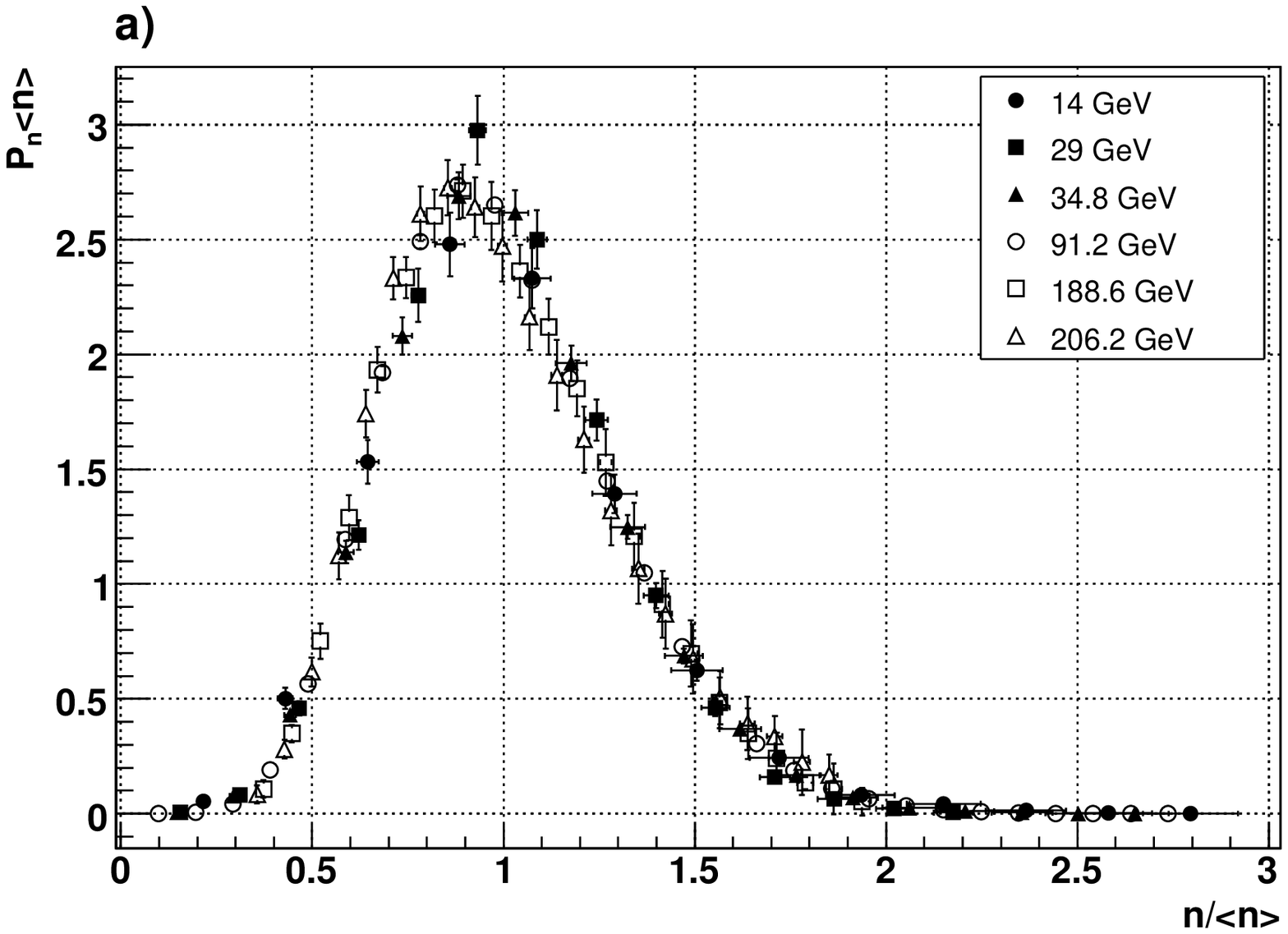}
\includegraphics[height=115mm,width=155mm]{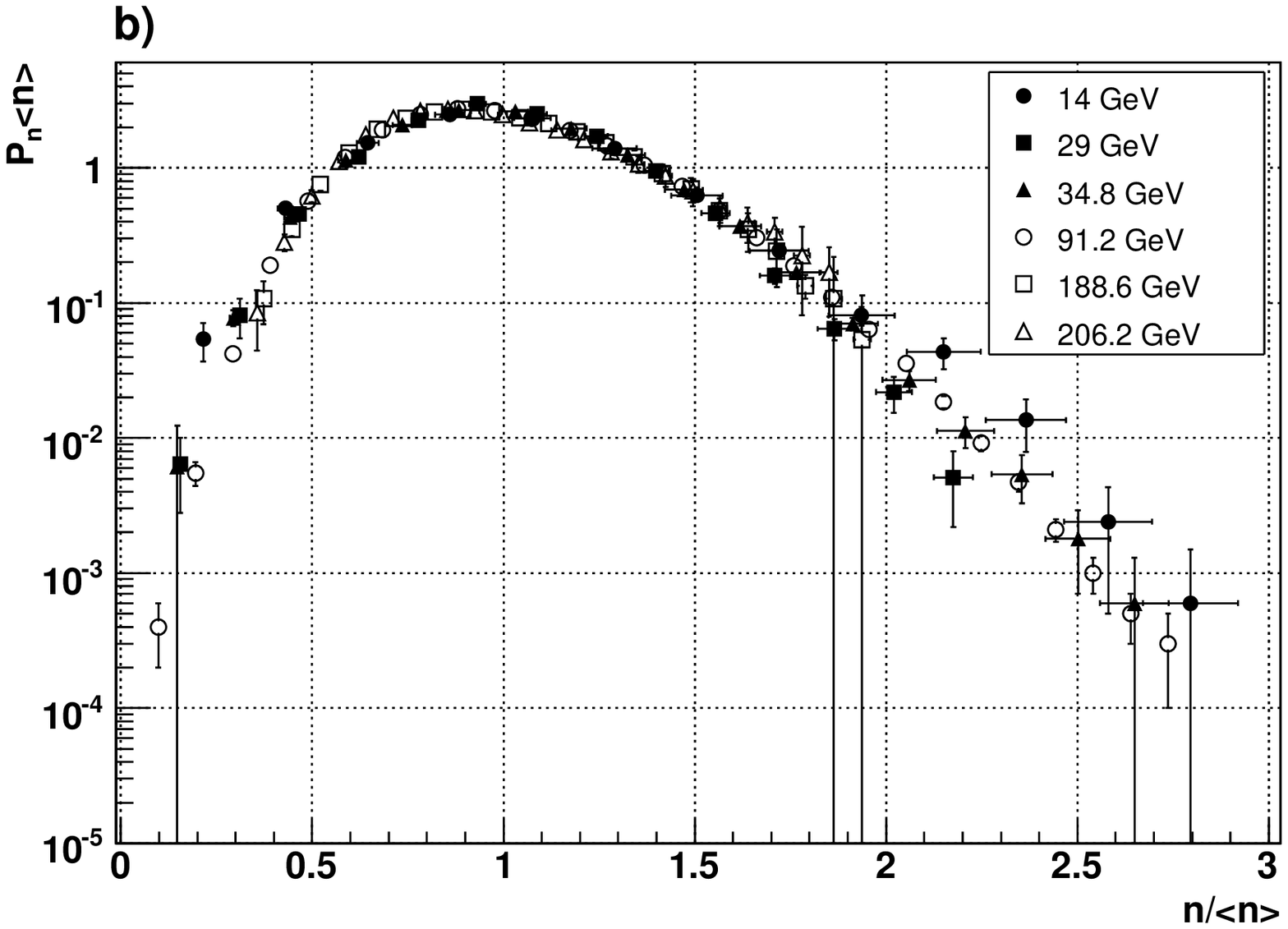}
\textbf{Fig.1.} The experimental KNO distribution. Both statical
and systematic errors are included
\end{figure}

\begin{figure}
\includegraphics[height=115mm,width=155mm]{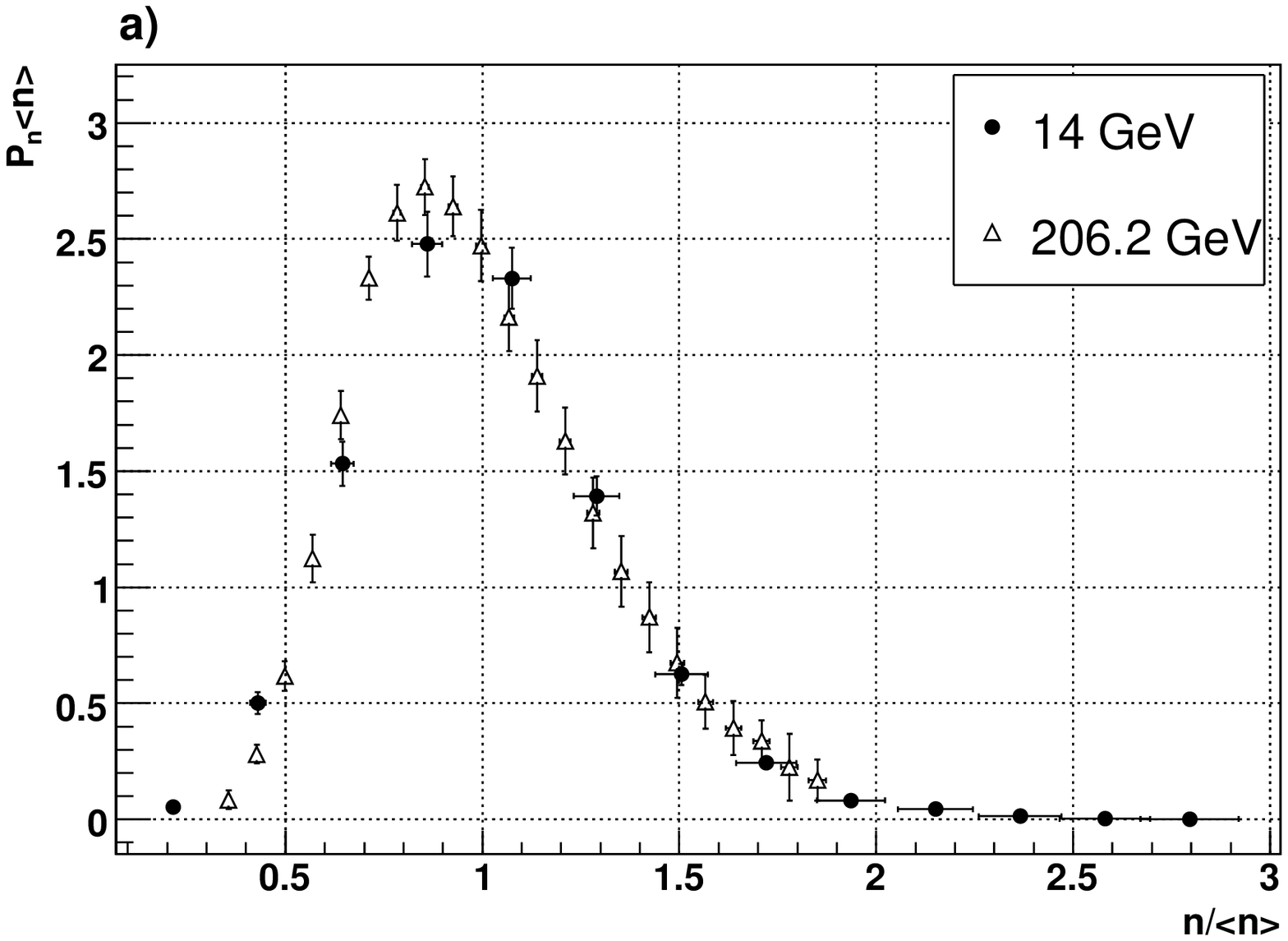}
\includegraphics[height=115mm,width=155mm]{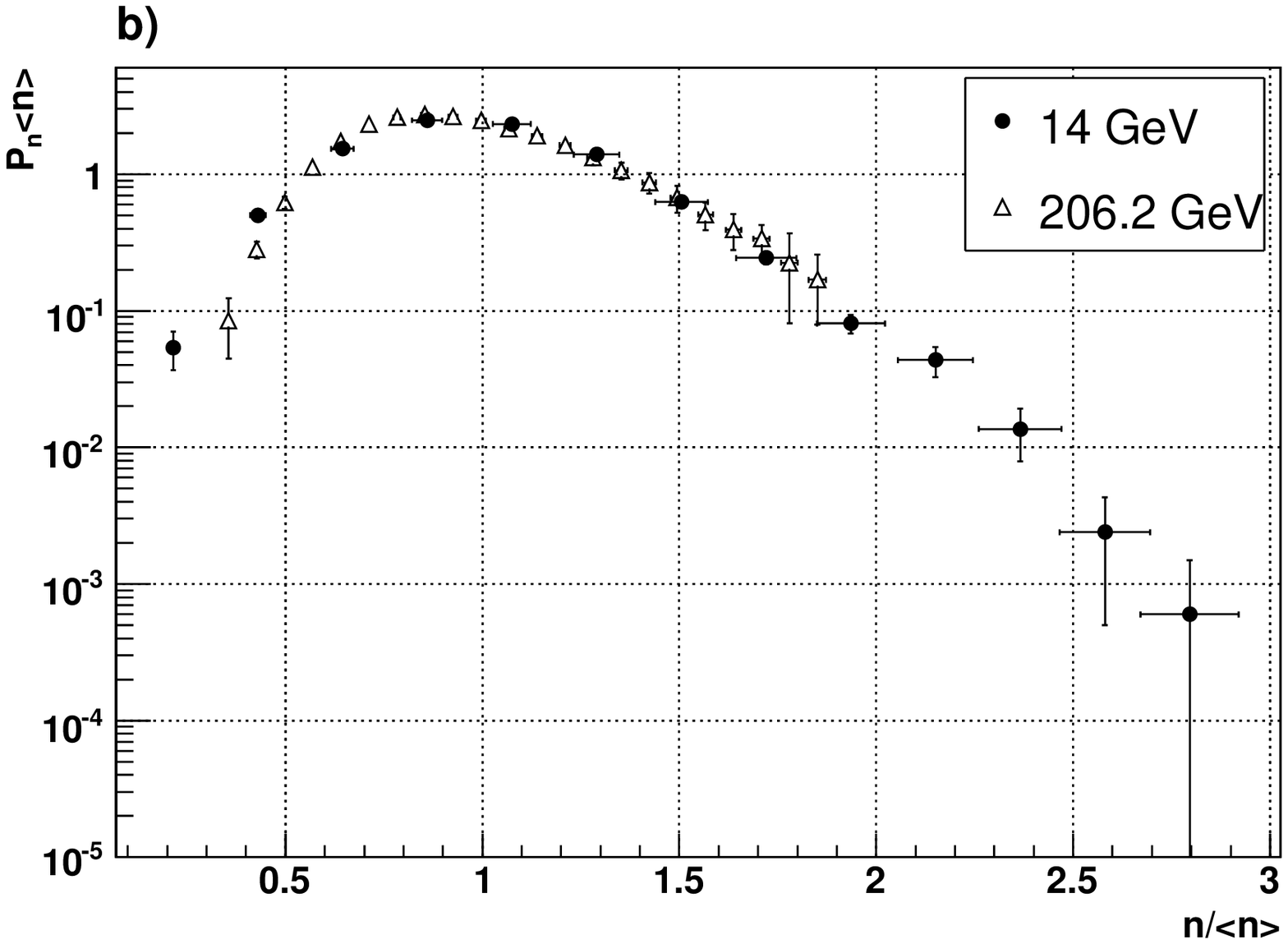}
\textbf{Fig.2.} The experimental KNO distribution. Both statical
and systematic errors are included
\end{figure}

\begin{figure}
\includegraphics[height=115mm,width=155mm]{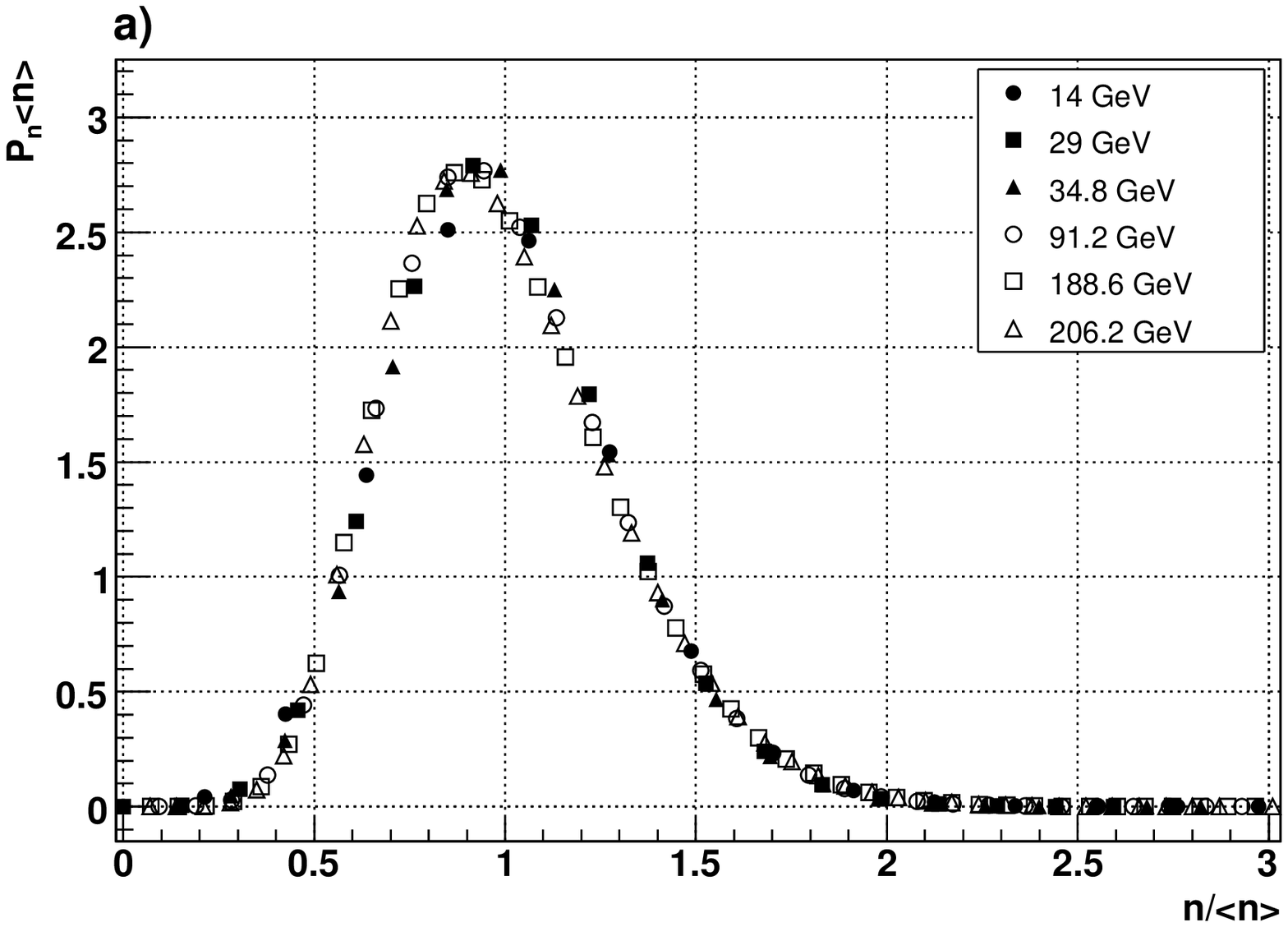}
\includegraphics[height=115mm,width=155mm]{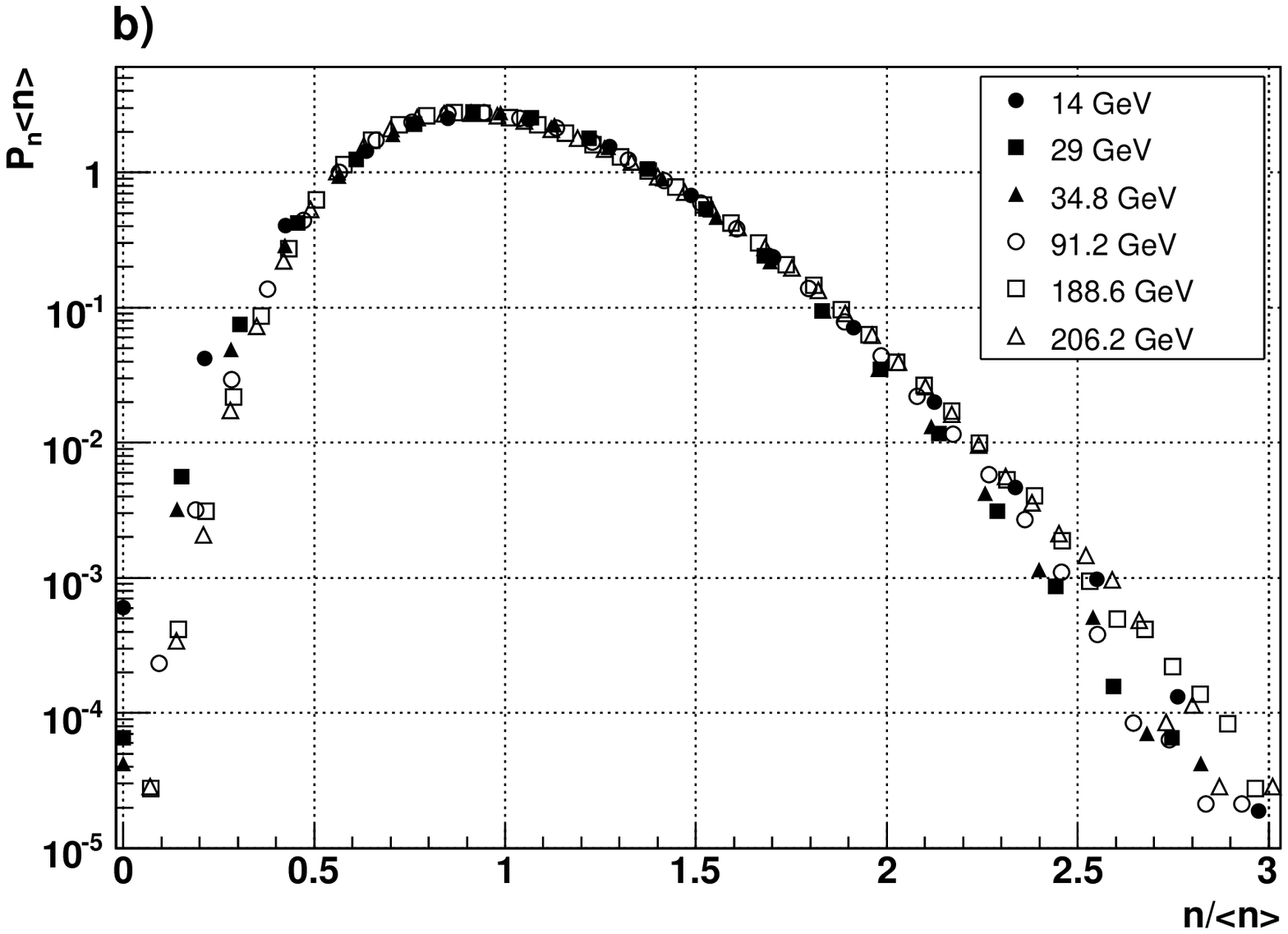}
\textbf{Fig.3.} KNO distribution generated by PYTHIA
\end{figure}

\begin{figure}
\begin{center}
\includegraphics[height=60mm,width=150mm]{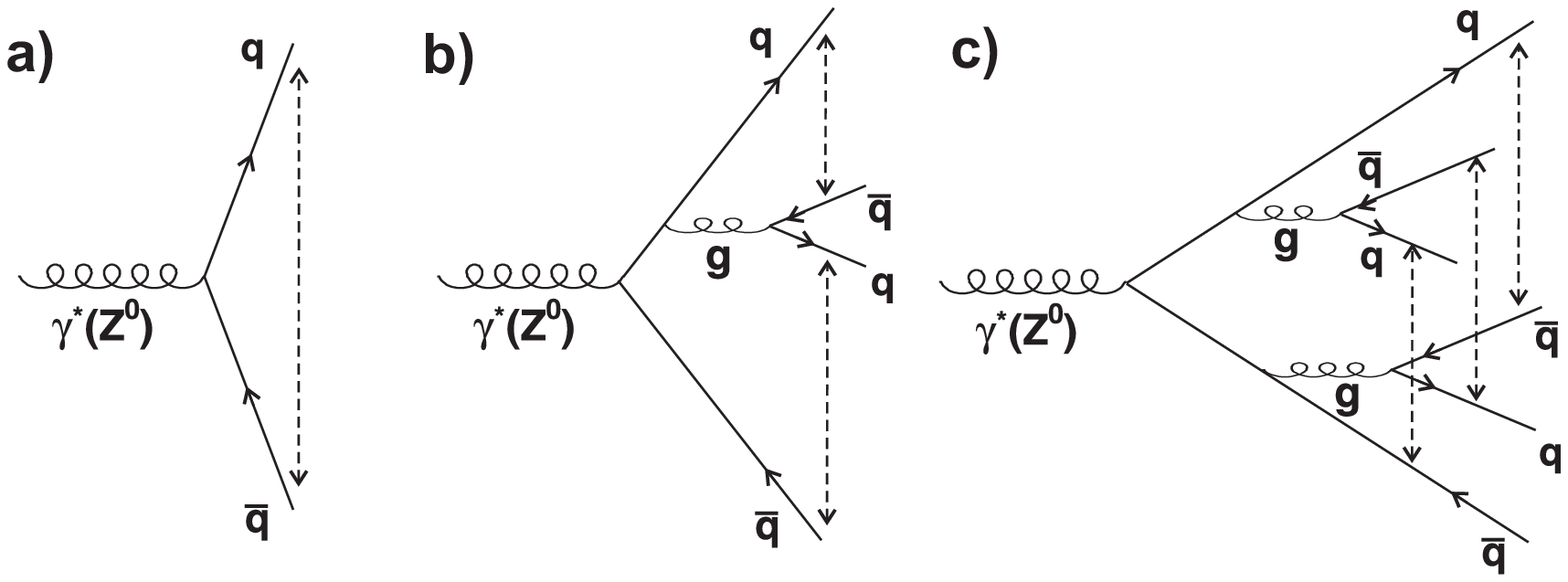}
\textbf{Fig.4.}
\end{center}
\end{figure}

\begin{figure}
\includegraphics[height=115mm,width=155mm]{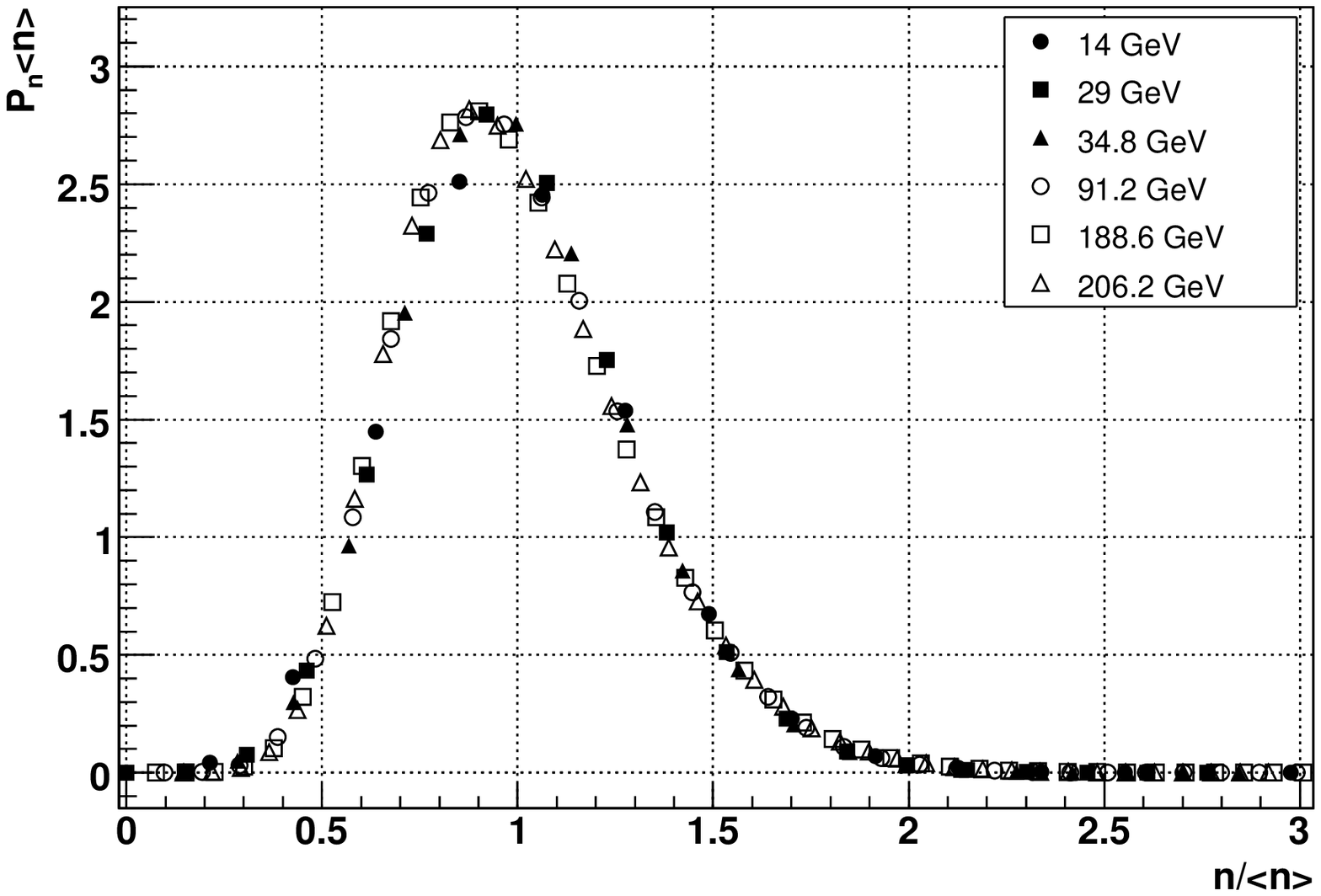}
\textbf{Fig.5.} The KNO distribution generated in PYTHIA for
events with on quark string
\end{figure}

\newpage
\begin{table}
\begin{tabular}{|c|c|c|c|}
\multicolumn{4}{l}{\textbf{Table 1.}}\\
\hline
\multicolumn{4}{|c|}{Experimental data}\\
\hline $\sqrt{s}$, GeV &$<\!\!n\!\!>$&$D$&$<\!\!n\!\!>/D$\\
\hline14&$9.30\pm0.41$&$3.07\pm0.28$&$3.03\pm0.31$\\
\hline29&$12.87\pm0.30$&$3.67\pm0.18$&$3.51\pm0.18$\\
\hline34.8&$13.59\pm0.05$&$4.14\pm0.39$&$3.28\pm0.33$\\
\hline91.2&$20.46\pm0.11$&$6.24\pm0.08$&$3.28\pm0.05$\\
\hline188.6&$26.84\pm0.32$&$7.89\pm0.22$&$3.40\pm0.10$\\
\hline206.2&$28.09\pm0.33$&$8.43\pm0.22$&$3.33\pm0.10$\\
\hline
\multicolumn{4}{|c|}{Data generated by PYTHIA}\\
\hline14&9.41&2.92&3.23\\
\hline29&13.11&3.84&3.41\\
\hline34.8&14.17&4.14&3.43\\
\hline91.2&21.16&6.28&3.37\\
\hline188.6&27.65&8.35&3.31\\
\hline206.2&28.56&8.65&3.30\\
\hline
\end{tabular}
\end{table}

\begin{table}
\begin{tabular}{|c|c|c|c|}
\multicolumn{4}{l}{\textbf{Table 2.}}\\
\hline $\sqrt{s}$, GeV&$C_3$&$C_4$&$C_5$\\\hline
\multicolumn{4}{|c|}{Experimental data}\\
\hline
14&$1.35\pm0.08$&$1.78\pm0.14$&$2.56\pm0.24$\\
\hline 29&$1.25\pm0.06$&$1.54\pm0.08$&$2.00\pm0.12$\\
\hline 34.8&$1.29\pm0.05$&$1.64\pm0.08$&$2.23\pm0.14$\\
\hline 91.2&$1.30\pm0.01$&$1.66\pm0.02$&$2.27\pm0.03$\\
\hline
188.6&$1.26\pm0.06$&$1.58\pm0.09$&$2.08\pm0.16$\\
\hline
206.2&$1.26\pm0.06$&$1.58\pm0.09$&$2.11\pm0.16$\\
\hline
\multicolumn{4}{|c|}{Data generated by PYTHIA}\\
\hline 14&1.30&1.66&2.25\\
\hline 29&1.27&1.58&2.10\\
\hline 34.8&1.27&1.58&2.09\\
\hline 91.2&1.28&1.61&2.17\\
\hline 188.6&1.29&1.65&2.26\\
\hline 206.2&1.29&1.65&2.26\\
\hline
\end{tabular}
\end{table}

\begin{table}
\begin{tabular}{|c|c|c|c|c|}
\multicolumn{4}{l}{\textbf{Table 3.}}\\
\hline &
\multicolumn{4}{|c|}{Number of strings}\\
\hline &1&2&3&4\\
\hline \multicolumn{5}{|c|}{$\sqrt{s}=14$ ÃýÂ}\\
\hline String length, $l$&3.75&5.52&&\\
\hline $<\!\!n\!\!>$&9.41&10.10&&\\
\hline Weight of event, \%&98.59&1.41&&\\
\hline
\multicolumn{5}{|c|}{$\sqrt{s}=29$ ÃýÂ}\\
\hline String length, $l$&4.83&6.70&8.99&\\
\hline $<\!\!n\!\!>$&13.03&14.62&15.81&\\
\hline Weight of event, \%&95.24&4.70&0.06&\\
\hline
\multicolumn{5}{|c|}{$\sqrt{s}=34.8$ ÃýÂ}\\
\hline String length, $l$&5.21&7.05&9.29&\\
\hline $<\!\!n\!\!>$&14.06&15.84&17.34&\\
\hline Weight of event, \%&94.08&5.81&0.11&\\
\hline
\multicolumn{5}{|c|}{$\sqrt{s}=91.2$ ÃýÂ}\\
\hline String length, $l$&6.66&8.87&11.59&14.02\\
\hline $<\!\!n\!\!>$&20.72&23.66&26.57&28.95\\
\hline Weight of event, \%&85.83&13.27&0.87&0.03\\
\hline
\multicolumn{5}{|c|}{$\sqrt{s}=188.6$ ÃýÂ}\\
\hline String length, $l$&8.09&10.63&13.64&16.55\\
\hline $<\!\!n\!\!>$&26.59&30.67&34.84&38.58\\
\hline Weight of event, \%&76.97&20.38&2.46&0.19\\
\hline
\multicolumn{5}{|c|}{$\sqrt{s}=206.2$ ÃýÂ}\\
\hline String length, $l$&8.24&10.85&13.89&17.04\\
\hline $<\!\!n\!\!>$&27.40&31.61&35.80&39.81\\
\hline Weight of event, \%&75.75&21.23&2.80&0.23\\
\hline
\end{tabular}
\end{table}


\begin{thebibliography}{9}
\bibitem{bib:1}
Z. Koba, H.\,B. Nielsen, P. Olesen, Nucl.~Phys. {\bf B40}, 317
(1972).
\bibitem{bib:2}
A.\,M. Polyakov, ZhETF {\bf 59}, 542 (1970).
\bibitem{bib:3}
J. Kogut, L. Suskird, Phys.~Rev. {\bf D10}, 732 (1974).

S. Brodsky, J. Gunion, Phys.~Rev.~Lett. {\bf 37}, 402 (1976).

E.\,G. Gurvich, Phys.~Lett. {\bf 87B}, 386 (1979).

A. Casher, H. Neuberger, S. Nussinov, Phys.~Rev. {\bf D20}, 179
(1979).
\bibitem{bib:4}
W. Braunschweig, R. Gerhards, F.J. Kirschfink et al., Z.~Phys.
{\bf C45}, (1989).
\bibitem{bib:5}
M. Derrick, K.\,K. Gan, P. Kooijman et al., Phys.~Rev. {\bf D34},
3304 (1986).
\bibitem{bib:6}
P. Achard, O. Adriani, M.Aguilar-Benitez et al.,
CERN-PH-EP/2004-024 (2004).
\bibitem{bib:7}
P.\,D. Acton, G. Alexander, J. Allison et al., Z. Phys. {\bf C53},
539 (1992).
\bibitem{bib:8}
P. Abreu, W. Adam, F. Adami et al., Z. Phys. {\bf C50}, 185
(1991).
\bibitem{bib:9}
P. Abreu, W. Adam, T. Adye et al., Eur.~Phys.~J. {\bf C18}, 203
(2000).
\bibitem{bib:10}
M.\,G. Bowler, P.\,N. Burrows, Z.~Phys. {\bf C31}, 327 (1986).
\bibitem{bib:11}
B. Andersson, G. Gustafson, G. Ingelman, T. Sjostrand, Phys.~Rep.
{\bf 97}, 31 (1983).
\bibitem{bib:12}
T. Sjostrand,  Phys.~Lett. {\bf 142B}, 420 (1984).

Nucl.~Phys. {\bf B248}, 469 (1984).
\bibitem{bib:13}
T. Sjostrand, S. Mrenna and P. Skands, JHEP05 (2006) 026 (LU TP
06-13, FERMILAB-PUB-06-052-CD-T) [hep-ph/0603175].
\bibitem{bib:14}
T. Sjostrand, FERMILAB-Pub-85/119-T (1985).
\bibitem{bib:15}
V.\,A. Abramovsky, O.\,V. Kancheli, Pis'ma~v~ZhETF, {\bf 31}, 566
(1980).
\end{thebibliography}
\end{document}